\documentstyle[11pt,paspconf,epsf]{article}

\begin{document}

\title{Photometric Redshifts in the CFDF}

\author{Mark Brodwin and Simon Lilly}
\affil{University of Toronto, 60 St. George St., Toronto, Canada M5S 3H8}

\author{David Crampton}
\affil{Dominion Astrophysical Observatory, 5071 West Saanich Rd., Victoria, Canada V8X 4M6}

\begin{abstract}
The Canada--France Deep Fields (CFDF) is a large, deep, multi-colour
imaging survey undertaken primarily at CFHT.  It is about 10 times
fainter than the CFRS (Lilly et al\ 1995a) and contains over 100 times
as many galaxies.  With three common fields, CFDF redshifts will be
estimated photometrically using the CFRS spectroscopic catalogue as a
training set.  The project will yield large numbers of galaxies and
will extend many of the CFRS results to higher redshifts and fainter
flux levels.
\end{abstract}



\section{Introduction}

Photometric redshifts are increasingly being used for science that
previously required expensive spectroscopic redshifts.  The
Canada--France Deep Fields (CFDF), a collaboration with LeF\`evre and
his Marseille group, is a large, deep multi--colour imaging survey
covering over 1 deg$^2$.  Photometric redshifts for of order $50,000$
galaxies in the CFDF will provide the means to address key issues in
galaxy formation and evolution.  The large areal coverage and
simultaneous depth of the CFDF will enable a measurement of the bright
end of the luminosity function (LF) (describing rare, extremely
luminous objects) to $z>3$, as well as an extension of the faint end
of the LF (ubiquitous faint objects) a factor of 10 deeper than
previous work (Lilly et al\ 1995b) for $z\leq1$.  The galaxy--galaxy
correlation function will be measured in the CFDF to $z\sim 1$ on
scales of order $10 h^{-1}$ Mpc, extending into the quasi-linear
regime.

\section{Observations}
The CFDF, undertaken on CFHT and NOAO facilities, is a deep,
multi--colour imaging survey that covers over 1 deg$^2$. Four
$30\arcmin \times 30\arcmin$ fields were imaged in \hbox{$U\!BV\!I$}\
to nominal $5\sigma\, (3\arcsec)$ sensitivities of $U_{AB}=26.00$,
$B_{AB}=25.75$, $V_{AB}=25.50$, and $I_{AB}=25.00$.  The image quality
is excellent, ranging from about 0.6\arcsec\ in $I$ to about
1.0\arcsec\ in $B$.  Several problems unique to CFHT's UH8k camera
(Metzger et al 1995) slowed the data reduction, however the pipeline
is now in place.  The reduction is about 50\% complete and should be
finished in a few months.

In order to obtain contiguous optical wavelength coverage (out to 1
micron), the original \hbox{$U\!BV\!I$}\ survey is currently being
supplemented with $R$-- and $Z$-- band data taken with the new CFH 12k
CCD mosaic camera at CFHT.  Some of this data is already in hand.  It
has been noted by several authors (e.g. Connolly et al 1997,
Fernandez-Soto et al 1999; see Yee 1998 for a review) that additional
NIR imaging is required to remove intrinsic degeneracies that exist
between distinct galaxy types at different redshifts.  KPNO time has
been allocated for infrared ($K'$) imaging of part of the CFDF fields
with IRIM.  The necessity of obtaining data at wavelengths longer than
the $I$--band was explicitly demonstrated in the detailed Monte Carlo
simulations that are described in the next section. The inclusion of
one IR band renders photometric redshifts immune to catastrophic
errors and reduces the dispersion at all redshifts.

\section{Photometric Redshifts}
Extensive Monte Carlo simulations were performed in order to test the
feasibility of photometric redshift estimation in the CFDF.  Repeated
interpolation between the Coleman, Wu, and Weedman (1980, hereafter
CWW) spectral energy distributions (SEDs) for E, Sbc, Scd, and Irr
galaxies resulted in 13 distinct template SEDs.  The CWW SEDs do not
extend blueward of 1400 \AA.  While strong L$\alpha$ emission aids
both in the detection of faint galaxies and in estimation of their
redshifts, it is often weak or absent due to severe dust obscuration.
In an effort to remain as conservative as possible, the CWW SEDs were
extrapolated to 1216 \AA\ and were set to zero blueward.  No modelling
of the line was attempted, and little would be gained by extrapolating
the SEDs right to 912 \AA.  The SEDs were convolved with generic
$\Delta \lambda/\lambda \sim 20\%$ filter response curves.  To create
the target galaxies, these SEDs were then redshifted into the range
$0\leq z \leq 2.6$ at $\Delta z = 0.1$ intervals and random
photometric errors consistent with those present in the CFDF were
added.  A $\chi^2$ fitting algorithm was used to compare the
redshifted templates with the target galaxies, with the minimum value
of $\chi^2$ corresponding to the best fitting redshift and template.
The simulation was run 100 times for each target galaxy (with
different random photometric error each time), enabling an estimation
of the typical dispersion in the best--fit redshift.

Several speakers at this meeting have pointed out that the use of
spectroscopic training sets result in better redshift estimates than
the use of templates derived from population synthesis models or
measurements of local SEDs (as in these simulations).  This is due in
part to the difficulty in accounting for galactic evolution, which is
naturally present in a spectroscopic catalogue.  However, in these
simulations no evolutionary {\em difference} between target and
template galaxies was present, therefore the accuracy should be
similar to what would be obtained on real galaxies using an empirical
training set.  In the final analysis the $\sim 600$ CFRS galaxies with
spectroscopic redshifts to $z\leq1.3$ will be used to train the
redshift--finding algorithm.
\begin{table}[!hpbt]
\begin{large}
\begin{center}
\begin{tabular}{|c||c|c|c|c||}\hline
Redshift&\multicolumn{4}{c||}{Simulated $\sigma_z$ with $U\!BV\!RI\!Z$}\\ \hline
Interval&~E/S0~&~~Sbc~~&~~Scd~~&~~Irr~~\\ \hline
0.25--0.75&0.13&0.09&0.08&0.09\\
0.75--1.25&0.16&0.15&0.10&0.10\\
1.25--1.75&0.16&0.17&0.36&0.21\\
1.75--2.25&0.05&0.42&0.64&0.17\\
\hline
\multicolumn{5}{c}{}\\ \hline
Redshift&\multicolumn{4}{c||}{Simulated $\sigma_z$ with $U\!BV\!RI\!Z\!K'$}\\ \hline
Interval&~E/S0~&~~Sbc~~&~~Scd~~&~~Irr~~\\ \hline
0.25--0.75&0.07&0.09&0.07&0.08\\
0.75--1.25&0.05&0.07&0.07&0.08\\
1.25--1.75&0.04&0.06&0.06&0.10\\
1.75--2.25&0.03&0.10&0.10&0.17\\ \hline
\end{tabular}
\end{center}
\end{large}
\caption{Summary of simulated redshift dispersions for faint
$I_{AB}=24$ galaxies imaged in \hbox{$U\!BV\!RI\!Z$} and
\hbox{$U\!BV\!RI\!Z\!K'$}\ in Monte Carlo simulations.  The
dispersions are averaged over $\Delta z=0.5$ redshift bins.  With the
optical data, the redshifts are good to $z\leq 1.25$, whereas the
addition of $K'$ allows accurate prediction at all redshifts}
\label{table}
\end{table}

\begin{figure}
\plotone{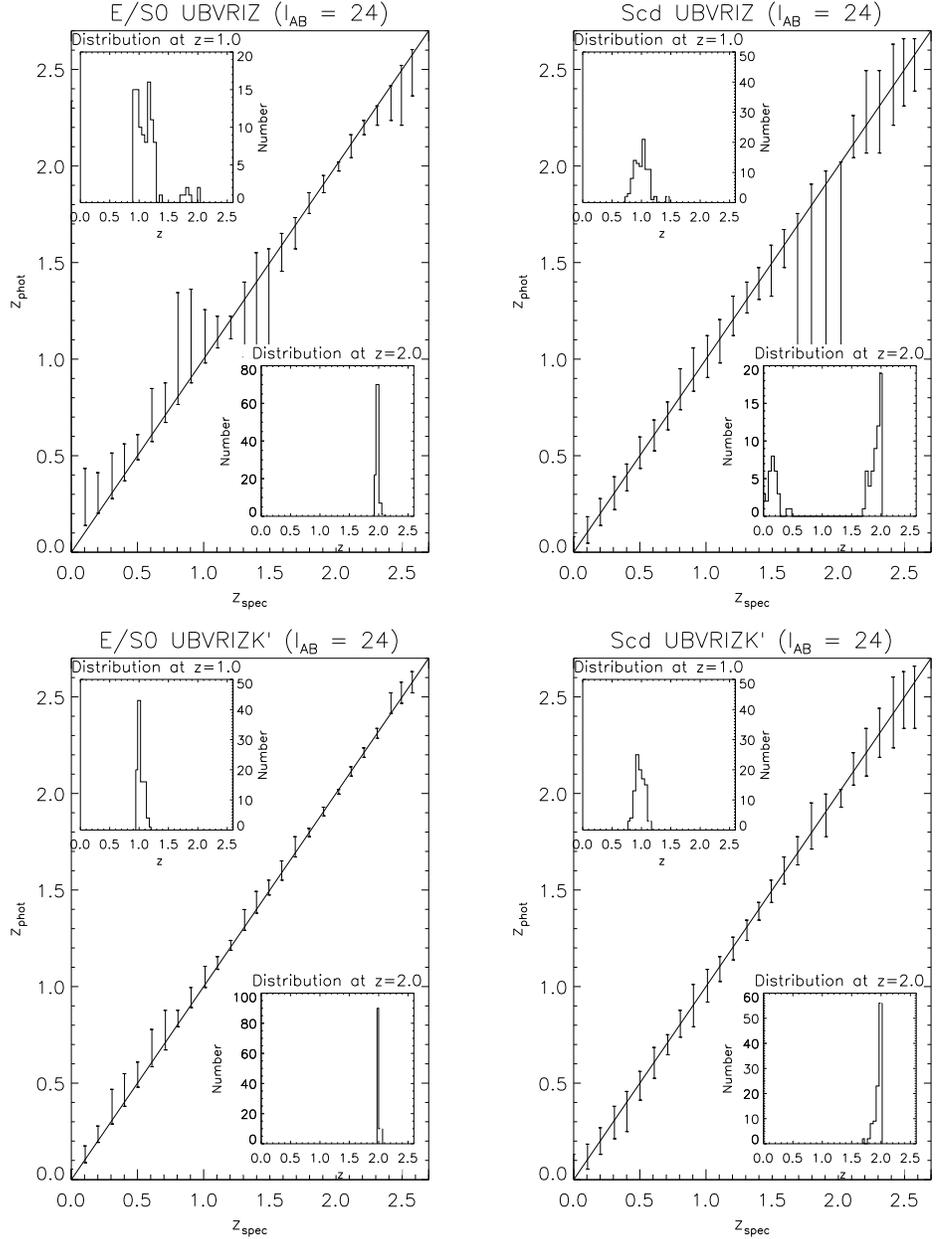}
\caption{The top panels plot $z_{phot}$ vs. $z_{spec}$ obtained
for early-- and late--type $I_{\mbox{{\small AB}}}=24$ model galaxies
for which \hbox{$U\!BV\!RI\!Z$} data (at the nominal sensitivities of
the CFDF) are available.  The full error bars span the central 2/3 of
the best-fit redshifts.  The insets of these panels show the complete
redshift distributions at $z=1$ (essentially gaussian) and $z=2$
(clearly demonstrating the occurrence of catastrophic errors).  The
bottom panels show the results obtained when $K'$ data is added.  The
error bars are reduced and the catastrophic errors (insets) are
eliminated at all redshifts.}
\label{z}
\end{figure}

The top panels of Figure \ref{z} illustrate the results for
$I_{AB}=24$ early-- and late--type galaxies with the
existing/scheduled \hbox{$U\!BV\!RI\!Z$}\ data. Aliasing between
($z\sim0$) ellipticals and ($z\sim2$) spirals is apparent. This is
likely due to the confusion of the Balmer break and the redshifted UV
cutoff.  Note that between $0.25 \leq z \leq 1.25$ the results are
unaffected by this degeneracy and the photometric redshifts for these
faint $I_{AB}=24$ galaxies have mean dispersions of about $\sigma_z
\leq 0.16$ (see Table \ref{table}).  The insets show the photometric
redshift distribution obtained in the 100 trials at $z=1$ and $z=2$.
The $z=1$ redshift distribution is basically free of catastrophic
errors in the redshift estimate, whereas the above--mentioned aliasing
produces a strongly non--gaussian redshift distribution for late--type
galaxies at $z \sim 2$.  A similar distribution exists for $z\sim 0$
ellipticals.  Catastrophic errors would clearly make an analysis
impossible at these redshifts.

The addition of $K'$--data (bottom panels) removes all degeneracies
and allows more accurate photometric redshifts to be estimated at $0
\leq z \leq 2.6$ and probably to higher redshifts due to the presence of the
Lyman break.  There are no catastrophic errors at $z=1$ or $z=2$; the
redshift distributions are almost delta functions centered on the
correct redshift.

\begin{figure}
\plotone{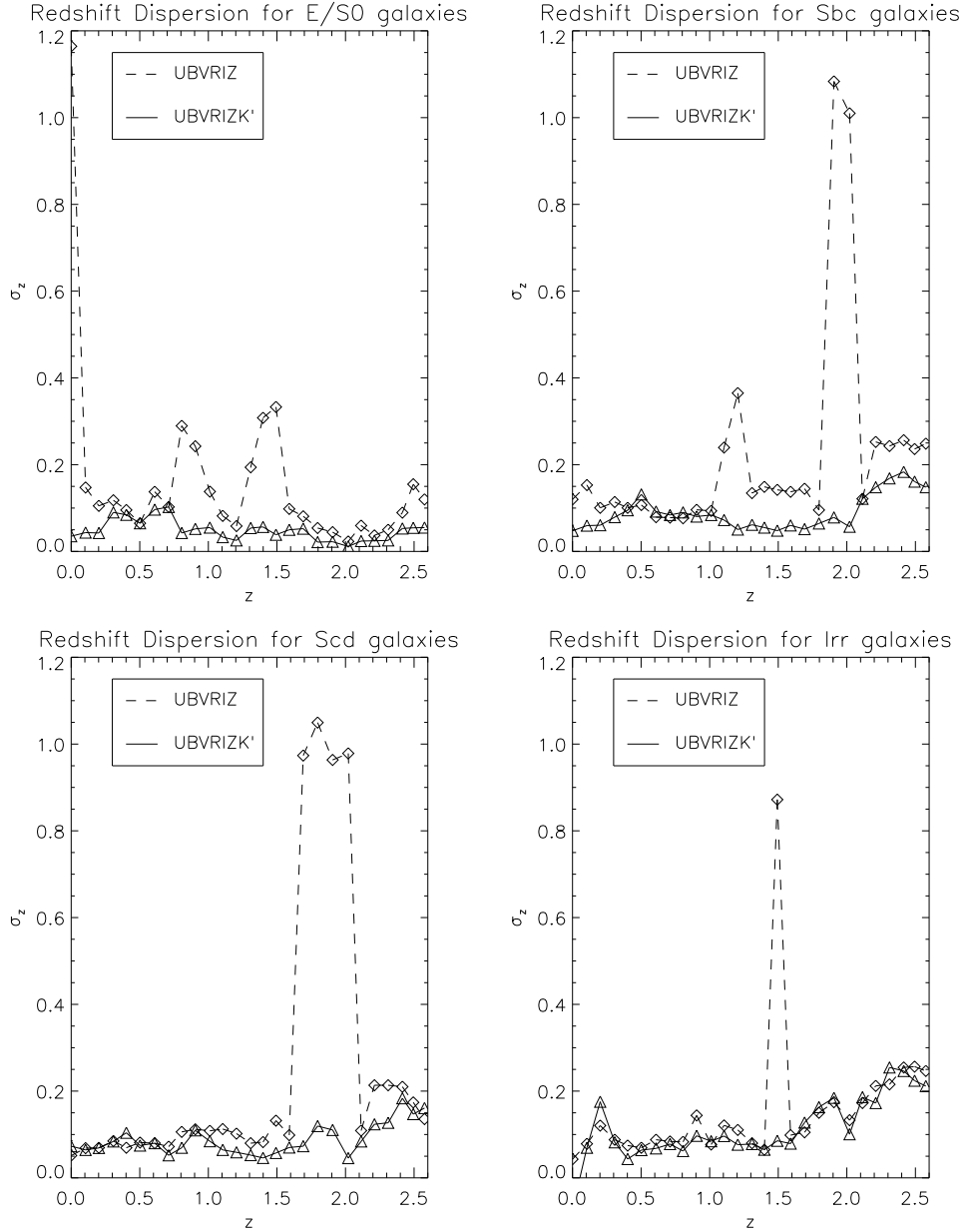}
\caption{Redshift Dispersion vs. Redshift is plotted for the CWW galaxy
types.  Between $0.2 \leq z \leq 1.2$, a mean dispersion of
$\overline{\sigma}_z \leq 0.16$ is obtained for all galaxies (see
summary in Table \ref{table}), with no contamination from other
redshift regimes.  The aliasing of high--$z$ spirals and low--$z$
ellipticals is readily apparent, but will not affect analyses at $z
\sim 1$.  The addition of $K'$--data removes this degeneracy and
allows accurate ($\overline{\sigma}_z \leq 0.1$) photometric redshifts
for all galaxy types to at least $z \sim2$ with no catastrophic errors
over the entire redshift range $0 \leq z \leq 2.6$.}
\label{sigma}
\end{figure}

Figure \ref{sigma} shows the redshift dispersion as a function of
redshift for each of the original CWW galaxy types for the optical and
optical+NIR filter sets.  The degeneracy between $z\sim0$ ellipticals
and $z\sim2$ spirals is clearly seen for the $U\!BV\!RI\!Z$ estimates,
but disappears when $K'$ is added.  (The error spike in $z\sim 1.5$
late--type galaxies is an artifact caused by the inclusion of negative
redshift template ellipticals used to obtain gaussian errors at
$z=0$.)  At $0.2 \leq z \leq 1.2$, the mean redshift dispersion for
all types of $I_{AB}=24$ galaxies is $\overline{\sigma}_z \leq 0.16$,
with {\em no contamination from higher or lower redshift objects}.
Thus analyses performed at $z \sim 1$ should be immune to catastrophic
redshift errors.  Note that with the addition of $K'$ data, the mean
dispersion drops to $\overline{\sigma}_z \leq 0.1$ for all galaxies to
at least $z \sim2$, with {\em no catastrophic errors at any
redshifts}.  The salient features of this Figure are summarized in
Table \ref{table}.

\section{Conclusions}

The CFDF is a large, deep, $U\!BV\!RI\!Z\!K'$ imaging survey covering
over 1 deg$^2$ and containing $\sim 100,000$ galaxies to
$I_{AB}=25$.  The science goals of the CFDF include measurements of
the luminosity and correlation functions of galaxies to $z \sim 1$.
As accurate photometric redshifts are required for these analyses,
detailed Monte Carlo simulations were conducted to characterize the
expected redshift errors.  These simulations indicate that for faint
$I_{AB}=24$ galaxies, redshift dispersions of $\overline{\sigma}_z
\leq 0.16$ can be expected in the range $0.2 \leq z \leq 1.2$ for all
galaxies imaged in $U\!BV\!RI\!Z$, with no significant occurrence of
catastrophic errors in redshift estimation.  For the data additionally
imaged in the near infrared ($K'$), the dispersion drops to
$\overline{\sigma}_z \leq 0.1$ over the range $0 \leq z \leq 2$
($\overline{\sigma}_z \leq 0.08$ at $z\sim1$) with no catastrophic
errors at these or higher redshifts.  It is therefore expected that
the redshift accuracy in the CFDF is sufficient to allow its science
goals to be achieved.

\acknowledgments
We'd like to thank the organizers for putting together this excellent
and timely workshop. M.\,B. would like to thank Stefan Mochnacki for
the use of his computer systems, as well as acknowledge a Reinhardt
Travel Award which partially funded the trip to Pasadena.

\end{document}